\documentclass[
  ,draft            
  ]
  {aipproc}

\layoutstyle{6x9}

\begin{document}

\title{Photon Production and Photon-Hadron Correlations in Relativistic Heavy Ion Collisions}


\author{Charles Gale}{
  address={Department of Physics, McGill University, 3600 rue University, Montreal, QC, Canada H3A 2T8}
}



\begin{abstract}
We calculate the production of real photons in relativistic nuclear collisions at RHIC, consistently with the quenching of fast partons. We go beyond one-body observables, and evaluate photon-triggered fragmentation functions, in the kinematical window corresponding to that of experimental measurements.
\end{abstract}

\maketitle


\section{Introduction}
The formation of a plasma of quarks and gluons has been a long-standing goal of contemporary nuclear science for a few decades now, and it is fair to write that this goal has been attained with the advent of RHIC (the Relativistic Heavy Ion Collider) at Brookhaven National Laboratory. Indeed, new physics has been discovered by the experimental program associated with this facility which now enters an exciting phase of characterization and of precision physics. In this context, real and virtual photons are penetrating probes as they suffer essentially no final state interaction. They constitute observables that are complementary to other hard probes, such as QCD jets.  We report here on calculations of jets and real photons at high $p_{T}$, and of photon-triggered hadron distributions in relativistic nuclear collisions at RHIC. 

\section{The evolving jets}
\begin{figure}[h]
\includegraphics[width=7cm]{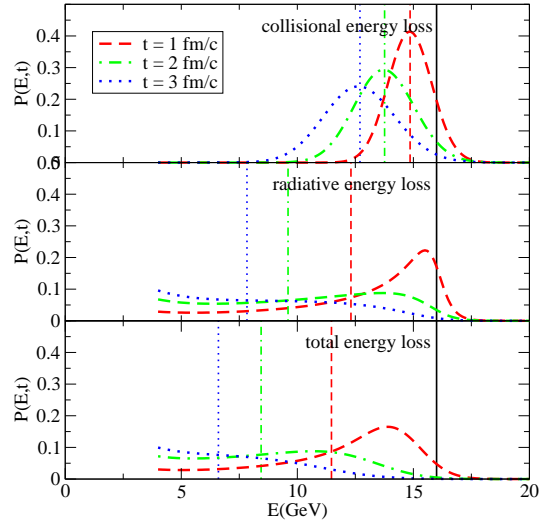}
\caption{A comparison of the effects of radiative and collisional energy loss on an initial quark with energy 16 GeV, traveling through a static medium at $T$ = 400 MeV. The number distribution at various times is plotted as a function of the energy. The initial state is a delta function at E = 16 GeV, and each vertical line corresponds to the average energy at the specified time. The calculation is from Ref. \cite{Qin:2007rn}. See however Ref. \cite{sgq} for an update bearing some qualitative differences.}
\label{rad-elastic}
\end{figure}

Arguably, one of the most sensational observations at RHIC has been the dramatic suppression  of high $p_{T}$ hadrons in central A+A collisions, compared with those measured in p+p events (multiplied by the number of binary collisions in a nucleus-nucleus event). This is quantified by the nuclear modification factor (in a given centrality range, which is correlated with the impact parameter, $b$): $R_{AA}^{h} (p_{T}, y) = 1/N_{\rm coll} (b) \left[ \left(d^{3} N^{h}_{AA} (b) / d^{2}p_{T} dy \right) / \left(d^{3}N_{pp}^{h} / d^{2} p_{T} dy\right)\right]$. At high transverse momentum, hadrons mostly originate from the fragmentation of QCD jets. However, before they escape the medium, hard partons will interact with the thermal medium. In the finite-temperature field theory approach utilized here \cite{Turbide:2005fk} the evolution of the parton distribution function $P(E,t) = d N(E, t)/dE$ can be modeled by a set of coupled Fokker-Planck equations generically written as 
\begin{eqnarray}
\frac{d P_{j} (E, t)}{d t} = \sum_{a, b} \int d \omega \left[P_{a} (E+ \omega, t) \frac{d \Gamma_{a \to j} }{d \omega d t}(E+ \omega, \omega) - P_{j} (E, t) \frac {d \Gamma_{j \to b} }{d \omega d t}(E, \omega) \right]
\end{eqnarray}
In the first term, a parton $j$ of energy $E$ is born from a parent $a$ of energy $E+\omega$. In the integral $- \infty < \omega < \infty$, which takes care of both energy loss and gain from the thermal medium.
The sum runs over different parton species and $d \Gamma_{j \to a} (E, \omega) / d \omega dt$ is the transition rate for the partonic process $j \to a$, calculated with the techniques of AMY \cite{AMY}. The hard partons may loose or gain energy by the radiation of gluons, or by elastic collisions with the hot medium. The relative importance of these two mechanisms is shown in Figure \ref{rad-elastic}.  

\section{Photon production}
There are many sources that produce photons in relativistic nuclear collisions, and they can be classified in two categories: on whether or not  they depend on the temperature of the medium. 
\subsection{Sources at $T = 0$}
Prompt photons, emitted during the very first instants of a nuclear collision, will represent an important background in the context of the search for signals of the quark-gluon plasma. Those in turn can be decomposed in two distinct sources: direct and fragmentation photons \cite{aur}. The direct photons are produced from the early hard collisions between partons in the nucleons of the projectile and target nuclei. At leading order in the strong coupling, the photon production proceeds via quark-antiquark annihilation ($q  \bar{q} \to g  \gamma$) and QCD Compton scattering ($q (\bar{q}) g \to q (\bar{q}) \gamma$). There is also a contribution to real photon from fragmenting hard QCD jets. This component is well-known for the case of, say, pp collisions \cite{aur}, but in nucleus-nucleus collisions the jet can propagate through the quark-gluon plasma, interact, and thus lose energy prior to its fragmentation. This introduces a non-trivial path and angle-dependence into the process of calculating the yield of photons produced through jet fragmentation \cite{Turbide:2005fk,Qin:2009bk}. 
\subsection{Sources at $T \neq 0$}
The most obvious source that requires a finite temperature treatment is that of photon production through the interaction of thermal components either from the quark-gluon plasma side of the QCD phase diagram, or from the hadrons in the confined sector. Emission rates for those have been established \cite{AMY,trg}. However, at hight $p_T$ thermal photons will not play an important role but it is nevertheless important to have their emission rate under control. Another thermal component is that of jet-medium photons \cite{fms,Turbide:2005fk}. These photons, owing to phase space considerations, will have the energy of the initial hard particle and their measurement is of primary importance as they represent an independent confirmation of the conditions for jet quenching \cite{Turbide:2005fk,Qin:2009bk}. Similary, a jet propagating in the hot and dense medium will produce electromagnetic radiation through bremsstrahlung. Finally, the jet fragmentation photons will now acquire a temperature dependence, through the energy loss mechanism and the coupling to the thermal medium. Note that all of the sources discussed here and above exists for dileptons, with appropriate adjustments \cite{dileps}. An importance milestone in such calculations is to first verify the correctness of photon spectra in p+p collisions, and then of those measured in nuclear collisions \cite{Qin:2009bk}. Importantly, the hydrodynamic evolution used here and in these cited works also yields a set of hadronic observables consistent with measurements.

\subsection{Photon-hadron correlation}

Going beyond one-body observables is important, as correlation studies will impose more stringent requirements on the underlying physics and might therefore highlight some theoretical differences that might otherwise have remain hidden. In this context, additional insight on jet quenching should be gained by triggering on a ``near side'' photon and measuring the distribution of charged hadrons in the opposite direction \cite{wang}. 
\begin{figure}[h]
\includegraphics[width=7cm]{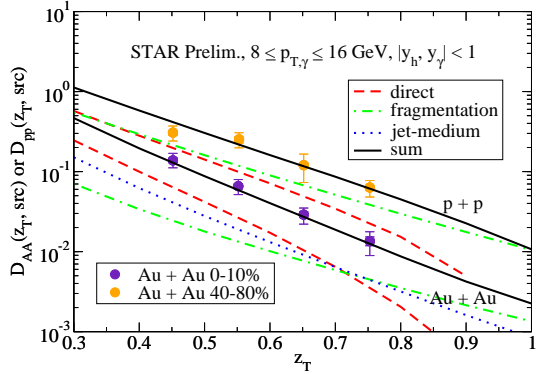}
\caption{The photon-tiggered fragmentation function in p+p and Au+Au collisions at RHIC. The different contribution are explained in the text. The data is from Ref. \cite{star}, and the calculation is from Ref. \cite{Qin:2009bk}.}
\label{D_AA}
\end{figure}
For this purpose, a useful variable is the photon-triggered fragmentation function: $D_{AA} (z_T,p_T^\gamma) = p_T^\gamma P_{AA} (p_T^h|p_T^\gamma)$, where $z_T = p_T^h/p_T^\gamma$, $P_{AA}$ is the yield per trigger: the momentum distribution of produced hadrons on the away side, given a trigger photon of momentum $p_T^\gamma$ on the near side. A calculation of $D_{AA}(z_T, p_T^\gamma)$ is shown in Figure \ref{D_AA}. Having a complete theory enables a breakdown of the contribution into its different components: At small $z_T$, roughly half the away-side hadrons are tagged by direct photons, while at higher values of $z_T$ a large amount of hadrons on the opposite side come from jets tagged by jet-medium photons and fragmentation photons. The agreement with the data is satisfying, and clearly the methods outlined here have great potential for future analyses as different parts of the available phase space will reveal different sources of photons. Finally, as the LHC will produce a plethora of jet events, the techniques briefly alluded to here will also be of great use in analyses there.


\begin{theacknowledgments}
I am grateful to all my collaborators for their help in the work presented here. 
  This research was funded in part by the Natural Sciences and Engineering Research Council of Canada. 
\end{theacknowledgments}

\end{document}